**TreeOTU: Operational Taxonomic Unit Classification Based on Phylogenetic Trees**

Dongying Wu*[1,2], Ladan Doroud[1], Jonathan A. Eisen[1]

[1]University of California, Davis, Davis, California 95616, USA
[2]DOE Joint Genome Institute, Walnut Creek, California 94598, USA

**Abstract**:

Our current understanding of the taxonomic and phylogenetic diversity of cellular organisms, especially the bacteria and archaea, is mostly based upon studies of sequences of the small-subunit rRNAs (ssu-rRNAs). To address the limitation of ssu-rRNA as a phylogenetic marker, such as copy number variation among organisms and complications introduced by horizontal gene transfer, convergent evolution, or evolution rate variations, we have identified protein-coding gene families as alternative Phylogenetic and Phylogenetic Ecology markers (PhyEco). Current nucleotide sequence similarity based Operational Taxonomic Unit (OTU) classification methods are not readily applicable to amino acid sequences of PhyEco markers. We report here the development of TreeOTU, a phylogenetic tree structure based OTU classification method that takes into account of differences in rates of evolution between taxa and between genes. OTU sets built by TreeOTU are more faithful to phylogenetic tree structures than sequence clustering (non phylogenetic) methods for ssu-rRNAs. OTUs built from phylogenetic trees of protein coding PhyEco markers are comparable to our current taxonomic classification at different levels. With the included OTU comparing tools, the TreeOTU is robust in phylogenetic referencing with different phylogenetic markers and trees.

**Introduction:**

Biologists have been using taxonomy to characterizing and understanding the living forms on earth for centuries. Not until the 1970s, Carl Woese pioneered the using the small subunit ribosomal rRNA sequences (ssu-rRNA) in phylogenetic studies of microorganisms that not only lead the discovery of the domain Archaea, but also revolutionized the field of taxonomic classification in terms of technology and methodology [1,2,3]. As powerful as ssu-rRNA sequences are in phylogenetic referencing, they have shortcomings such as complications introduced by horizontal gene transfer and convergent evolution [4,5,6], difficulties for aligning for some lineages [7], as well as the variation in copy numbers among different organisms that complicates abundance estimation [8,9]. To compensate the disadvantage of phylogenetic studies depending only on ssu-rRNAs, we've identified protein coding Phylogenetic and Phylogenetic Ecology markers (PhyEco) at different taxonomic groups, including 40 markers that cover both bacteria and archaea domains, and 114 bacteria specific markers [10].

Currently, there are two different ways to classify organisms into taxonomic groups. One depends on the sequence collections in current databases and adopting their taxonomic annotations as standards. Sequences from the environment are taxonomically classified according to their sequence similarities to the reference sequences in the databases. Such approaches are popularized in ssu-rRNAs taxonomic classification by projects such as greengenes [11], SILVA [12] and RDP [13]. Another taxonomic classification method is to divide the sequences into different groups called Operational Taxonomic Units (OTUs) based upon similarities between the sequences. Such approach is tremendously useful in biodiversity studies [14]. This approach has been exemplified by the ssu-rRNA sequence analysis packages QIIME [15] and MOTHUR [16]. There are many shortcoming of sequence similarity based OTU classification: first, sequences similarities sometimes don't reflect phylogenies [17,18,19,20]; second, sequence similarities cannot be calculated for partial sequences from metagenomics data that don't overlap [21]; third, different clustering algorithms generate OTU clusters with different granularities for the same set of sequences that are not related to phylogenies (e.g. the nearest neighbor, furthest neighbor and average neighbor clustering algorithms implemented in the MOTHUR package [16] ).

In 2011, Sharpton et. al. addressed some of the issues in their ssu-rRNA OTU classification pipeline PhyOTU by using phylogenetic distances from a tree instead of sequence similarities between sequences [21]. They've successfully solved the problem of the lack of distance measurements between non-overlapping sequences, and opened up the possibilities of using phylogenetic methodologies (such as corrections to convert measures of similarity to evolutionary distances) to improve the accuracy of OTU building [21]. Relying on the same set of clustering algorithm for OTU grouping as MOTHUR [16], PhyOTU is still have cluster granularity variation issues that are not related to phylogenies of the sequences [21]. Phylogenetic trees have been used for decades for taxonomic classifications, but the application falls into two categories. One is to assign taxonomies to sequences based on its position in phylogenetic trees with the current taxonomic classification of their neighbors as references.

The automated phylogenetic tree-based ssu-rRNA taxonomy and alignment Pipeline STAP [20] and genome and metagenome phylogenetic analysis package PhyloSift [22] are two examples of such an approach. The taxonomic classification of the two packages are fully automated but have problems of accurately assigning taxonomies to novel sequences with their close relatives absent from the reference databases. Another way to use phylogenetic trees for taxonomic studies is to identify taxonomic groups by manual examination [7]. The results of manual phylogenetic tree studies are subjected to different scientists' different interpretations.

The accelerated increase of computing power that is available to the research communities and the vast improvement of phylogenetic building speed have made phylogenetic tree building for unprecedented number of taxa a reality. For example, Fasttree have built a phylogenetic tree for 399,817 ssu-rRNA sequences [23]. Such advances provide great opportunities for us to carry out phylogenetic analysis for large number of sequences, but they make the traditional visualization dependent manual phylogenetic studies no longer practical.

We've developed an automatic phylogenetic tree base OTU classification program TreeOTU. Because TreeOTU only depends on phylogenetic trees, it simplifies OTU classification based on different PhyEco markers. To compare different OTU sets from different trees, we've also include in the TreeOTU package OTU comparing scripts (see methods). Our tree-base OTU classification and comparison algorithms can be fully automated, thus enables phylogenetic studies at a much larger scale than before. TreeOTU package and all the alignments and trees used in this study can be download from github and figshare.

**Discussion**

<u>TreeOTU addresses the issue of different rates of evolution of different lineages in a phylogenetic tree</u>

One major challenge for automatic phylogenetic OTU classification is the variation of evolution rates between different lineages [24]. The phylogenetic trees we build have very different depths for different clades because of evolution rate variation. If we define OTUs as taxa groups that share different common ancestors at a given time in history, the phylogenetic distances (PDs) of such ancestors from the root should be different among different OTUs in a phylogenetic tree. If we applied an arbitrary phylogenetic distance (PD) cutoff from the root to divide the taxa to OTUs, we usually break the fast evolving lineages into different OTUs, while force the slow evolving lineages into the same OTU even if they diverged very early in the history of time.

Taking evolution rate variation into account, we invented an approach to classify taxonomic groups using phylogenetic trees as references. Our program TreeOTU includes two critical steps. The first step involves the adjustment the depth of the node (node distance from the root) by normalizing it with the average of the depths of all its child leaves (leaf distance from the root). In reality, phylogenetic trees are usually biased with the dominations of certain lineages because if either sampling bias [25] or true biodiversity reality in the environment [26,27]. To limit the dominant organisms' impact on the average distance calculation, we weight the average of leave depths by their phylogenetic contributions (see methods), thus thousands of identical organisms are counted only once.

The adjusted node depth calculations depend on all the child leaves of the nodes, thus often resulting in larger depths (distance from the root) of parent nodes than their child nodes. In addition, phylogenetic trees involving metagenomic data usually have artificial long branches because of short sequence length and bad alignments [7] that would have large negative effects on node depth calculation. The challenge leads to our second critical step in TreeOTU: a iterative strategy of node evaluation and OTU picking from the leaves towards the root (see method). The iterative strategy not only prevents an already classified OTU from interfering other taxa's grouping, but also effectively quarantines artificial long branched taxa.

TreeOTU takes into account both tree topologies and branch lengths. But yet, we are not expecting all the OTU groups to be monophyletic in the phylogenetic tree. The main reason is what we called "the leftover effect": for a clade in a tree, some of its taxa was classified as distinct OTU groups, while the remaining taxa still satisfy our criteria to form an OTU that is not monophyletic in the tree. We demonstrate "the leftover effect" in Figure 4: 10 OTUs from the genus *Nocardia* built by TreeOTU from "the All-species Living Tree" [28] are mapped on the tree from which the OTUs are classified. 6 out of the 10 OTU groups are monophyletic, while 4 groups are not (Figure 4, OTU T1, T5, T8 and T9). "The leftover effect" is inherent in our iterative OTU classification approach, and it is not necessarily a disadvantage. The scenarios of

"the leftover effects" that are demonstrated in Figure 4 occur in manual phylogenetic tree analysis that often require scientists' judgment calls.

The advantage of tree-based OTU classification over sequence similarity based methods

Large amount of researches have demonstrated that sequence similarity based taxonomic studies sometimes disagree with phylogenetic analysis [17,18,19,20]. TreeOTU provides a great opportunity to quantify the difference of the two approaches. We use the bacterial and archaeal ssu-rRNA alignment and the tree from the All-species Living Tree project [28] as an example for the comparison. We used MOTHUR [16] to generate OTUs from the alignment and compared the results with OTUs generated by TreeOTU based on the maximum likelihood tree built by RaxML from the same alignment [28,29]. We used adjusted mutual information (AMI) to quantify the similarity of OTU sets built from the two methods [30]. If a set of MOTHUR OTUs is identical to a set of tree-base OTUs, the AMI value is 1. The more different the two sets are, the smaller AMI value gets. Two drastically different OTU sets give a AMI value close to 0 [30].

We've compared all the pairs of OTU sets between MOTHUR and TreeOTU at different cutoffs, and demonstrate the results in Figure 3. We observed good matches between MOTHUR OTUs and tree-bases OTUs, and the corresponding sequence similarity cutoffs for MOTHUR and TreeOTU cutoffs correlate very well. In Figure 3, out of the 145 OTU sets from different TreeOTU cutoffs are included, 134 can find at least one similar MOTHUR OTU set with the AMI value larger than 0.8. However, the differences between the two different OTU classification methods are substantial. We observed no identical OTU sets between the two methods with the exception of the ones that separate the two domains of bacteria and archaea (data not shown in Figure 3). There are only 8 tree-based OTU sets in Figure 3 that can find the best MOTHUR OTU matches with the AMI values larger than 0.9.

To highlight the difference between MOTHUR and TreeOTU, we used the example of MOTHUR OTUs with sequence similarity distance of 0.03, a cutoff used by many to define species separation [31]. We compared the MOTHUR OTU set with its closest tree-based OTUs that is obtained at a TreeOTU cutoff of 0.04 with an AMI value of 0.81 (Figure 4). MOTHUR groups 9133 ssu-rRNA sequences into 4830 OTUs. Out of the 1375 MOTHUR OTUs that contain more than one ssu-rRNA sequences, we've identified 733 with identical OTUs by TreeOTU at the cutoff of 0.04. There are two main reasons that 642 non-singleton MOTHUR OTUs cannot be perfectly matched by TreeOTU at the optimum cutoff. First, one TreeOTU cutoff might split MOTHUR OTUs that agrees with phylogenetic tree topology. Indeed, 299 of MOTHUR OTUs can find perfect matches by TreeOTU at cutoffs ranging from 0.03 to 0.05 other than 0.04. The second reason is that sequencing similarity based OTUs sometimes don't accurately reflect phylogeny, thus disagree with phylogenetic trees. Figure 4 demonstrates one of such example. In one clade within the genus of Nocardia in the ssu-rRNA All-species Living Tree [28], we have 8 MOTHUR OTUs with multiple members (Figure 4) with 7 of them cannot be captured by TreeOTU at any cutoffs (M1, M2, M3, M4, M5, M7, M8). In many cases, MOTHUR and TreeOTU disagree but both are in line with the phylogenetic tree topology (for example: M2, M5, M7 and M8). While other differences simply cannot be resolved and directly reflect the fundamental disagreement between sequence similarity and phylogenetic based methodologies (for example: M1, M3 and M4).

Our study demonstrates that tree-based OTUs and sequence similarity based OTUs are comparable but different. TreeOTU generates OTUs that are always more faithful to phylogenetic trees in case of any differences between the two different taxonomic classification methodologies (Figure 4).

## TreeOTU results are comparable to our current taxonomic structure

We retrieved the taxonomies of the ssu-rRNA sequences in the All-species Living Tree from the NCBI taxonomic database, and build OTU sets at the following 6 levels: species, genus, family, order, class and phylum. The six NCBI taxonomy based OTU sets are compared with TreeOTU results at different cutoffs to identify the cutoffs that are correspondent to different NCBI taxonomic levels (Figure 3). NCBI taxonomies at different level correspond to different TreeOTU cutoffs very well, with higher taxonomic level correlated with OTU sets with higher cutoffs. We don't show the AMIs at the species level because they are extremely low across all the TreeOTU cutoffs. The All-species Living Tree project includes one sequence from one species [28], thus the species level OTUs gives very low AMI value when compare to similar TreeOTU results. The reason is that the random chance noise is extremely high in mutual information based similarity comparisons due to the extremely high granularities of the species level OTU sets [30].

We were curious to know if we can observe similar correlations between our current taxonomic structure and OTUs built from phylogenetic trees of protein coding Phylogenetic and Phylogenetic Ecology markers (PhyEco) [10]. We've built phylogenetic trees for ssu-rRNA, and 40 PhyEco markers from the bacteria and archaea genomes in the IMG database [10,32]. In addition, we concatenated 38 alignments of PhyEco markers to build a "genome tree" for the comparison. We compared the OTUs from all 42 phylogenetic trees at different TreeOTU cutoffs against the taxonomic assignments at 6 levels in IMG (species, genus, family, order, class and phylum) [32]. The AMI values of all the comparisons are illustrated in Figure 5. Similar to the correlations between the OTUs from the All-species Living Tree and NCBI taxonomy (Figure 3), from the species level all the way up to the class level, the TreeOTU cutoffs that give the best fitting OTU sets rise accordingly for all the phylogenetic trees we've examine. All the 42 phylogenetic trees give similar phylogenetic history that is well captured by the current taxonomic classifications in the IMG database [32] . Different trees have different TreeOTU cutoffs that yield the best fitting OTUs to the same taxonomic level because different genes have different evolving rate and patterns even they share similar evolving history. An extreme example in our study is the PhyEco marker BA00026 (ribosomal protein S12/S23). The phylogenetic tree of BA00026 have long edges between nodes closer to the root, thus the optimum TreeOTU cutoffs are all vastly smaller than all the rest 41 trees at all the taxonomic levels in this study (Figure 5).

The comparisons between the OTUs built by TreeOTU and the IMG taxonomic classification at the phylum level yield low AMIs for almost all the trees with the exception of a few including ssu-rRNA and the "genome tree" built from concatenated 38 PhyEco markers (Figure 5). We used F measurement to quantify the support that each phylum gets from the tree based OTUs [33]. The F1 score that measures the similarity between two OTU groups is between 0 and 1(see method) [33]: If two OTUs are identical, F1 score is 1; the more different the two OTUs are, the lower F1 score gets.  All genomes from one phylum (e.g. *Proteobacteria*) form the query group, and it was compared against each one of the OTU group generated by TreeOTU at different cutoffs. The F1 score of the best matching tree based OTU from a tree represents the support the phylum gets from the tree. We've examined 20 phyla (2 from Archaea and 18 from Bacteria), and the top F1 scores are illustrated in Figure 6.  If a maximum F1 score larger than 0.8 is an indication of good support from the tree, Bacteria *Planctomycetes* as a group are not well supported by 22 out of the 42 trees (Figure 6). The following phyla are not well supported by at least 10 trees: Bacteria *Tenericutes*, *Chloroflexi*, *Proteobacteria*, *Verrucomicrobia* and Archaea *Euryarchaeota*.  These phyla are the main reasons of the low AMI values in comparisons between the tree-based OTUs and IMG phylum classification (Figure 5). If we hold current IMG phylum classification as the gold standard [32],  ssu-rRNA is the best tree to represent it: the ssu-rRNA tree not only strongly supports all the phyla we examined (Figure 6), but also gives the top matching OTU set with IMG phylum classification with an AMI value of 0.91 among all 42 trees (Figure 5).

Tree-based OTU classification provide a robust way to compare different phylogenetic trees

It has been established that taxonomic classification of one gene can be misleading because of lateral gene transfers and convergent evolution [4,5,6]. In addition to the traditional ssu-rRNA as phylogenetic markers, we've identified protein coding genes that can be used as Phylogenetic and Phylogenetic Ecology markers (PhyEco) [10]. However, comparing taxonomic results based upon phylogenetic trees of different genes has not been robust and usually requires visualization and manual examination [7,34]. TreeOTU transforms tree structure to cluster data structure, thus provide a novel way for phylogenetic tree comparison that can be automated.

The direct comparison of two trees is accomplished by build tree-based OTUs at different TreeOTU cutoffs for both tree, and use adjusted mutual information (AMI) [30] to measure the similarity between all pairs of the OTU sets. This approach is exactly like the comparison of tree-base OTUs against MOTHUR OTUs at different cutoffs (Figure 3). The AMI matrixes not only can be visualized (Figure 3), but also can be used for other analysis such as correlation studies using sliding windows to compare two trees at different phylogenetic levels that can be automated.

Another way to compare different phylogenetic trees is to compare OTUs from different trees against one OTU sets we called "standard OTUs". Standard OTUs can be OTUs from one phylogenetic tree at a certain TreeOTU cutoff, and they can also be a group of taxa such as current NCBI taxonomy classification to serve as a standard to test hypothesis. This approach has been demonstrated by the example of comparing IMG genome taxonomy with TreeOTU results from 42 phylogenetic trees (Figure 5, 6). The example highlights the importance of calibration in tree-based OTU comparison. One TreeOTU cutoff for one tree cannot be applied to other trees, all the OTUs at different TreeOTU cutoffs from different trees need to be compared. In addition, we have to emphasize the importance of tree rooting, because different rooting strategy results in drastically different comparison results. In all the studies we described in this manuscript, all the trees were rooted mid-point between the archaea and bacteria clades. For huge phylogenetic trees, rooting a tree using graphic interface tools such as Dendroscope [35] is simply impractical. We've developed a series of scripts that root trees from the linux command line and they are available for download in the TreeOTU package (see Conclusion).

**Conclusion:**

In conclusion, we've developed the software package TreeOTU that classifies OTU according to phylogenetic tree structure and branch lengths. We've demonstrated that OTU sets built by TreeOTU at different cutoffs correlate with current classification for bacteria and archaea at different taxonomic levels. TreeOTU builds OTUs that are also comparable to sequence similarity based OTU classification methods, while tree-based OTUs are more faithful to phylogenetic tree topologies. By transforming phylogenetic tree structures into data sets of OTU clusters, TreeOTU opens up new possibilities of comparing evolving histories of different genes. We are in the process to exploring the novel applications of TreeOTU in the fields of metagenomics diversity studies [34,36] and gene function referencing by phylogenetic profiling [37].

The TreeOTU package includes OTU classification, comparison and tree rooting scripts, as well as the alignments, trees and NCBI/IMG taxonomic classification information related to this research. It can be downloaded from github (https://github.com/dongyingwu/TreeOTU) and figshare (http://figshare.com/articles/TreeOTU/783077).

**Methods:**

Measurement of the position of a node in a rooted phylogenetic tree

The normalized distance of a node to the leaves of its sub-tree is used to measure the position of a node in a rooted phylogenetic tree, we call it PN (Position of a Node). PN of node n in a rooted tree is defined by equation (1): $Rn$ is the distance of the node n to the tree root, $Dn$ is the distance of the node to its child leaves. $Dn$ is defined by equation (2): $Di$ is the distance between leaf i and node n, $Pi$ is the independent phylogenetic contribution of leaf i to the sub-tree from the node n. $Di$ and $Pi$ are defined in equation (3) and (4): $Vi$ is the length of the edge connects leaf i to its parent node, m is a node between leaf i and the node n, $Vm$ is the length of the edge connects node m and its parent, $Cm$ is the number of child leaves from node m. An example of PN calculation is illustrated in Figure 1.

$$PN = \frac{Dn}{Dn + Rn} \quad (1) \qquad Dn = \frac{\sum (Pi \times Di)}{\sum Pi} \quad (2)$$

$$Di = Vi + \sum Vm \quad (3) \qquad Pi = Vi + \sum \frac{Vm}{Cm} \quad (4)$$

Algorithm of OTU identification based on the positions of the nodes in a phylogenetic tree

For a given edge in a rooted phylogenetic tree, the PN values (Position of the Node) of the nodes at both end of the edge are calculated. If one end of the edge is a leaf, the PN value of the leaf is set to 0. If the PN value of the node close to the root is larger than or equal to a given cutoff while the PN value of the node at the other end is not, the edge is cut and all the child leaves from the edge are group into one OTU group. If the PN values of both ends of an edge are lower than the cutoff, the edge is left intact. To prevent the scenario that PN values of both ends of an edge are all above the cutoff, a sequence of edge processing is followed to prioritize the evaluation and OTU grouping (Figure 2). At the very beginning, all the leaves in the tree are marked "CURRENT" which mean the edges that connect to them are under the first round of evaluation. Because all the leaves have PN values of zero, all edges connected to parent nodes above the cutoff are cut, and the released leaves are clusters into OTUs that each contains only one member. After all the edges connected to the leaves are processed, nodes with all its leaves gone are remove from the tree, and the leftover leaves are marked "PROCESSED". From this point, an iterative process is followed: first, all the nodes with all their child nodes/leaves marked "PROCESSED" are identified and labeled "CURRENT"; second, PN values of the parent nodes of "CURRENT" nodes are calculated. If the PN value of the parent node exceed the cutoff, the edge is cut and all the child leaves are grouped into one OTU group; third, the remaining "CURRENT" nodes are labeled "PROCESSED" and all the nodes with no child leaves are remove from the tree for the next iteration (Figure 2). After all the nodes in the tree have been processed, the leaves are grouped into a set of OTUs.

Adjusted mutual information based OTU cluster comparison

Adjusted mutual information (AMI) is used to quantify the similarity between different sets of OTUs. The measurement is based on Fred and Jain's normalized mutual information [38] with the adjustment for chance [30].

For a set of N taxa, two different sets of OTUs are generated: set U has R OTUs and set V has S OTUs. H(U) is the entropy of OTU set U, H(V) is the entropy of OTU set V, H(U,V) is the joint entropy of U and V. The entropies are defined by equation 5-7. In equation 5 and 6, $a_i$ is the number of taxa in OTU i; in equation 7, $n_{ij}$ is the number of taxa that are shared by OTU i in U and OTU j in V. I(U,V) is the mutual information between U and V, the measurement is defined by equation 8. Adjusted mutual information (AMI) between U and V is calculated by equation 9.

$$H(U) = -\sum_{1}^{R}(\frac{a_i}{N} \times \log_2 \frac{a_i}{N}) \quad (5) \qquad H(V) = -\sum_{1}^{S}(\frac{a_i}{N} \times \log_2 \frac{a_i}{N}) \quad (6)$$

$$H(U,V) = -\sum_{1}^{R}\sum_{1}^{S}(\frac{n_{ij}}{N} \times \log_2 \frac{n_{ij}}{N}) \quad (7) \qquad I(U,V) = H(U) + H(V) - H(U,V) \quad (8)$$

$$AMI(U,V) = \frac{2 \times (I(U,V) - E)}{H(U) + H(V) - 2 \times E} \quad (9)$$

In equation 9, E, the expected mutual information between U and V, is the average of mutual information values of 100 pairs of random OTU sets. Each random pair of OTU sets are generated by the permutation model [39, p214]: one have the same number of OTUs and taxa in each OTU as U, and the other have the same number of OTUs and taxa in each OTU as V, the mutual information between the two random OTU sets is calculated according to equation 8. AMI value is between 0 and 1: AMI of two identical sets of OTUs is 1; the more different the two OTU sets are, the smaller AMI value gets.

Quantify how one group of taxa is represented by an OTU set

To evaluate how well an OTU set captures a given group of taxa, F-measure of the group of interest against all the OTUs in the set is adapted [33]. For each comparison of the query group of taxa (Q) and an OTU (O), the F1 score is calculated by the equation 10: The precision *p* is the number of overlapping taxa between Q and O divided by the number of taxa in the OTU (O); The recall *r* is the number of overlapping taxa between Q and O divided by the number of taxa in the query group (Q).

$$F1 = 2 \times \frac{p \times r}{p + r} \quad (10) \qquad AF_{max} = \frac{F_{max} - E_{max}}{1 - E_{max}} \quad (11)$$

A series of F1 scores are obtained between the query group and all of the OTUs in the set, we use the maximum F1 score ($F_{max}$) to measure the OTU set's performance to capture the query group of taxa Q. To adjust for chance, $E_{max}$, the expected value of $F_{max}$ is calculated. The permutation model is used to generate random sets of OTUs [39, p214]. The taxa in the OTU set is shuffled and the randomized OTU set has the same number of OTUs and each OTU has the same number of taxa as the original OTU set. $F_{max}$ is calculated for the random set against the query group Q. $E_{max}$ is the $F_{max}$ average of the query group Q against 100 random OTU sets. The adjusted $F_{max}$ ($AF_{max}$) is defined by equation 11. If query group Q finds an identical OTU in one OTU set, $AF_{max}$ is 1. The more different between the best matching OTU in the OTU set and query group Q, the smaller $AF_{max}$ gets. $AF_{max}$ is set to 0.0001 if equation gives a value less than 0.0001.

Phylogenetic tree building for tree-base OTU classification
Genome sequence data were downloaded from Integrated Microbial Genomes (IMG) version 4 on January 4, 2013 [32]. The database include 16,182,124 peptide sequences from 4518 bacterial genomes and 462,347 peptide sequences from 182 archaeal genomes, as well as 14,047 ssu-rRNA sequences from 4330 bacterial genomes and 297 ssu-rRNA sequences from 180 archaeal genomes.

The following protocol was followed to select one copy of ssu-rRNA from each genome: Step 1, a alignment was built from 8608 non-redundant ssu-rRNA from IMG that were longer than 800nt by SINA aligner [40]; Step 2, a neighbor-joint tree was built by Fasttree [23], and the tree was rooted in the middle of archaea and bacteria domains; Step 3, OTUs was built by TreeOTU at a cutoff of 0.009, the longest copy of the ssu-rRNA genes was chosen to represent its genome if all the copies of ssu-rRNA from the genome were in one OTU group, thus genomes with ssu-rRNA genes spreading across multiple OTU groups were excluded. As a result, our IMG ssu-rRNA dataset in this study include 161 archaeal and 3630 bacterial genomes. An alignment was built for IMG ssu-rRNA sequence datasets by SINA aligner [40] and a maximum likelihood tree was build by Fasttree [23].

40 protein coding PhyEco markers that span both bacteria and archaea [10] are included in this study. Sequences were retrieved by HMMSEARCH using the Hidden Markov model (HMM) of the markers [10] against the IMG database [32] and aligned by HMMALIGN from the HMMER3 package [41]. A neighbor-joint tree was built for each gene family by Fasttree [23], and the tree was rooted mid point between the archaea and bacteria domain. For each IMG PhyEco gene family, OTUs were built by TreeOTU at a cutoff of 0. If a genome has members in the same PhyEco gene family that spread across multiple OTUs, the genome was excluded from that PhyEco gene family. Otherwise, one gene from one genome was selected for each PhyEco gene family. The numbers of genomes that are included in all the 40 PhyEco families are listed in table S1. All 40 single-copied PhyEco marker families were aligned by HMMALIGN [41]. 38 alignments out of the 40 families were concatenated: BA00035 and BA00038 were excluded because they cover less genomes than the rest, table S1; Only IMG genomes that have no less than 34 PhyEco markers were included in the concatenated alignment. Fasttree was used to build maximum likelihood trees with the WAG model for all 40 PhyEco genome families as well as the concatenated alignment [23].

Comparison of TreeOTU and MOTHUR

To compare OTUs built by TreeOTU with those by sequence similarity-based method, The ssu-rRNA maximum likelihood tree build by RaxML [29] as well as the alignment from the "The All-species Living Tree" Project (LTP) [28] was downloaded from SILVA website [12]. (http://www.arb-silva.de/projects/living-tree/, release 111, January 22, 2013). The LTP dataset includes 359 archaeal and 9335 bacterial ssu-rRNA sequences, only 9133 ssu-rRNA sequences with NCBI classification at any of the follow taxonomic levels are included in the study: genus, family, order, class and phylum.

MOTHUR was used to classify sequence similarity based OTUs from the alignment: OTU sets were built by the default furthest neighbor algorithm at sequence similarity distance cutoffs from 0 to 0.295, at an interval of 0.005 [16]. Meanwhile, OTUs were built by TreeOTU from the "All-species Living Tree" at cutoffs from 0 to 0.725 at an interval of 0.005. Adjusted mutual information (AMI) values were calculated between all pairs of tree-based OTU sets and MOTHUR OTU sets and the matrix is displayed in Figure 3.

**Figures:**

**Figure 1.** An example of PN (Position of Node) calculation. We use a rooted tree with 4 leaves as an example. The PN value of node n is obtained through the following calculation: node n has three leaves A, B and C. The phylogenetic distance of leaf A to node n ($D_A$) is 1.0+1.2=2.2 (equation 3); The independent phylogenetic contribution of leaf A to the sub tree of node n ($P_A$) is 1.0+1.2/2=1.6 (equation 4), because node m have two leaves, the independent contribution of A to the edge between nodes n and m is only half of the edge length. Similarly, for leaf B, $D_B$=3.2, $P_B$=2.6; for leaf C, $D_C$=3.0, $P_C$=3.0. The distance of the node n to its leaves is the average of all the distances of the leaves to node n weighted by their independent phylogenetic contributions to the sub tree, which amounts to 2.9 (equation 2). The distance from node n to the root is 1.8, thus the distance of all the child leaves of node n to the root is 1.8+2.9=4.7. The PN of node n is the distance of the node n to its child leaves (2.9) normalized by the distance of the root to its child leaves (4.7), which equals 0.62 (equation 1).

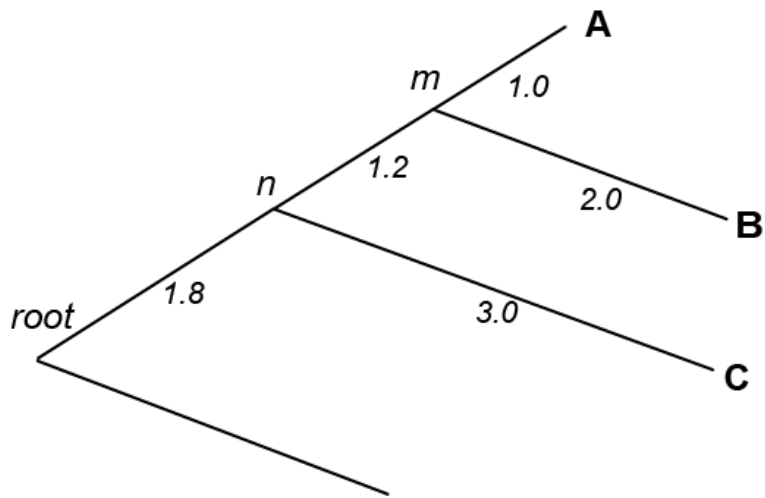

**Figure 2.** The protocol for OTU (operational taxonomic unit) identification from a phylogenetic tree.

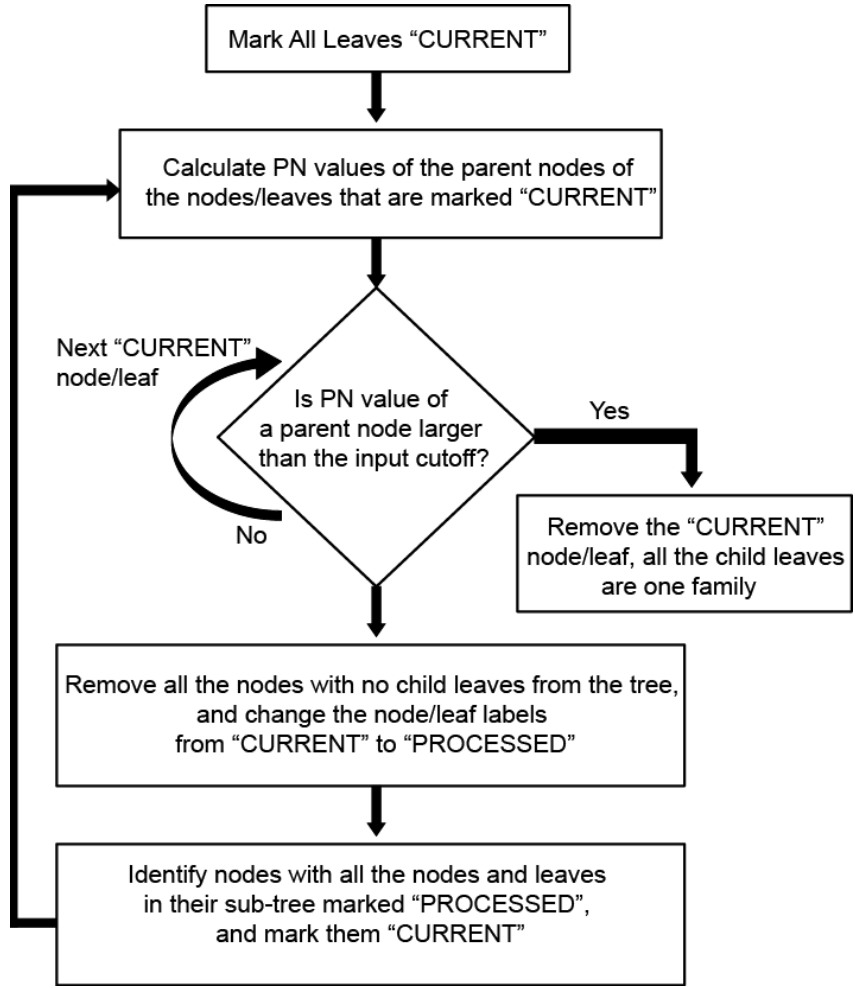

**Figure 3.** Adjusted mutual information values (AMI) between the OTUs built by TreeOTU and MOTHUR for the bacteria and archaea dataset from the "The All-species Living Tree" Project (LTP).

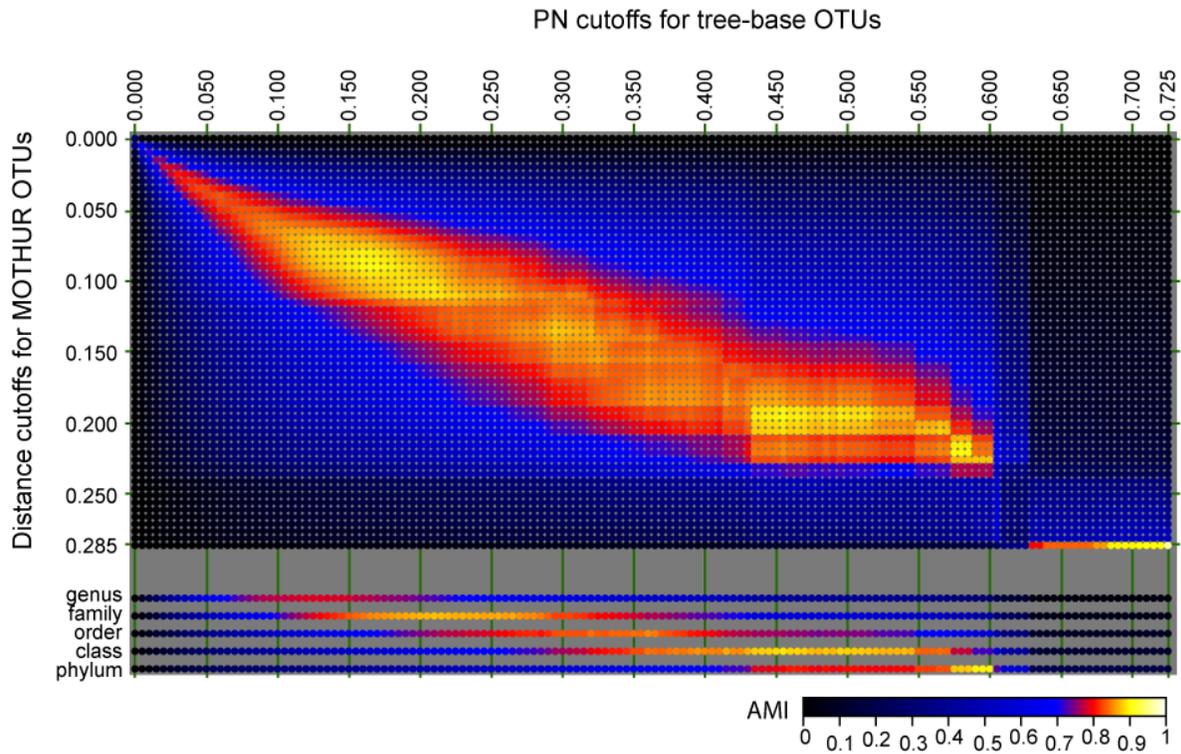

**Figure 4.** Comparison of TreeOTU and MOTHUR in a clade within the genus *Nocardia* in the "All-species Living Tree". 10 Tree based OTUs are generated at a TreeOTU cutoff of 0.04 (T1-T10), and they are labeled with colored text. 8 MOTHUR OTUs are built at a sequence similarity distance of 0.03 (M1-M8), and they are labeled with white text shaded by colored boxes.

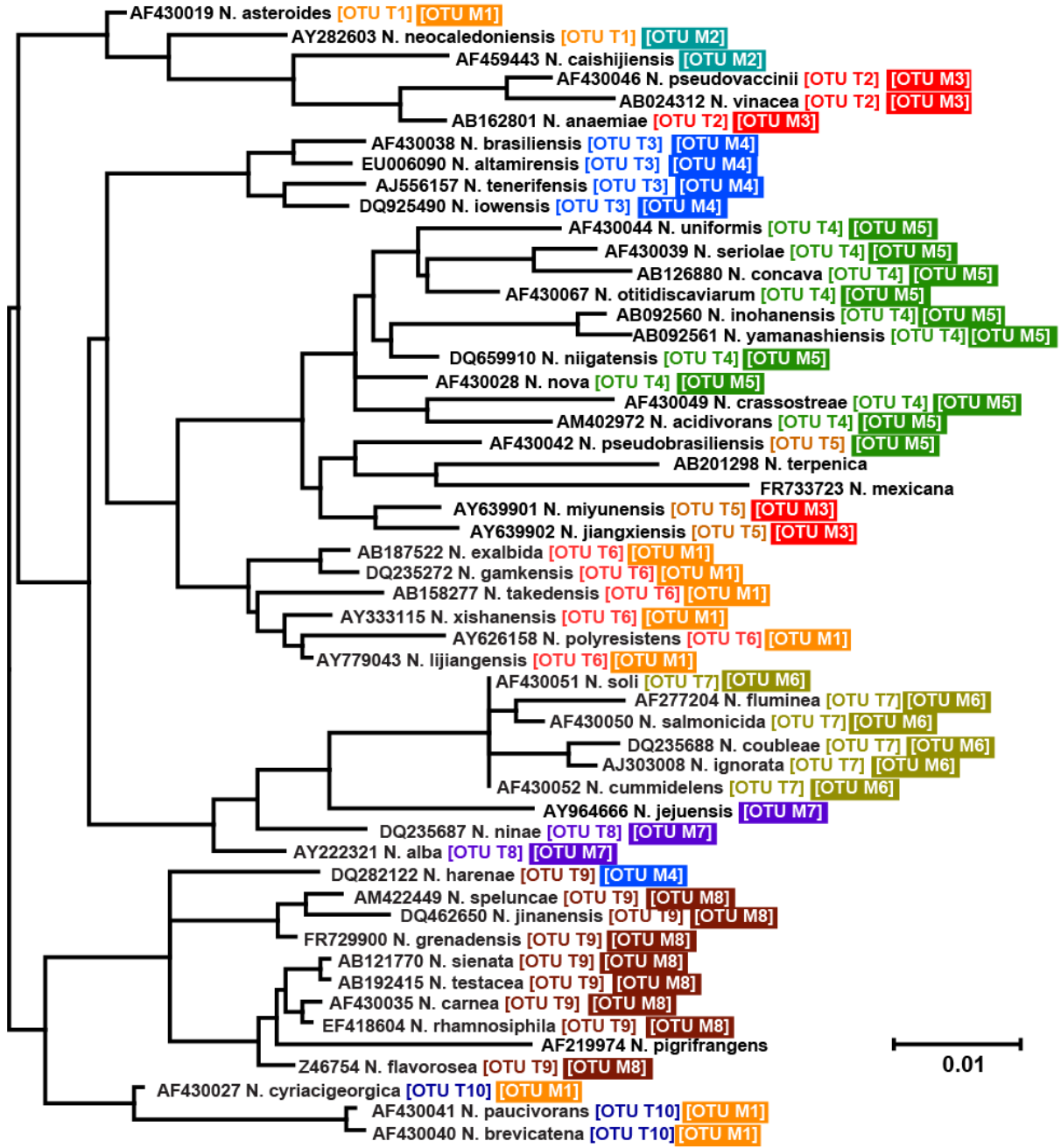

**Figure 5.** Comparison of IMG taxonomies and OTUs built by TreeOTU at 6 taxonomic levels (species, genus, family, order, class and phylum) from the phylogenetic trees of ssu-rRNA, 40 PhyEco markers and the concatenated alignment 38 PhyEco markers. The adjusted mutual information (AMI) values for the 42 trees are listed in the same order as the level of family in all the other taxonomic levels.

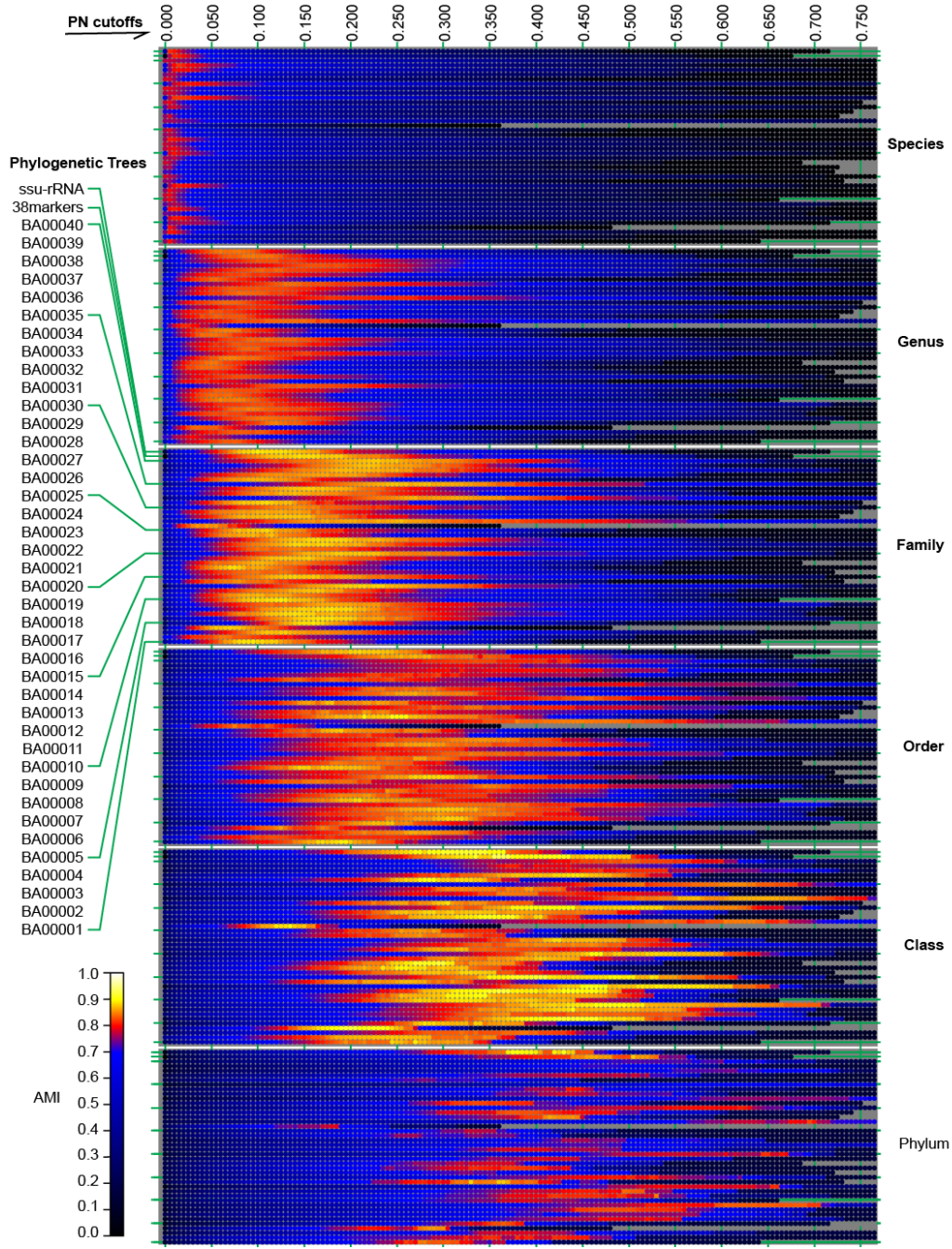

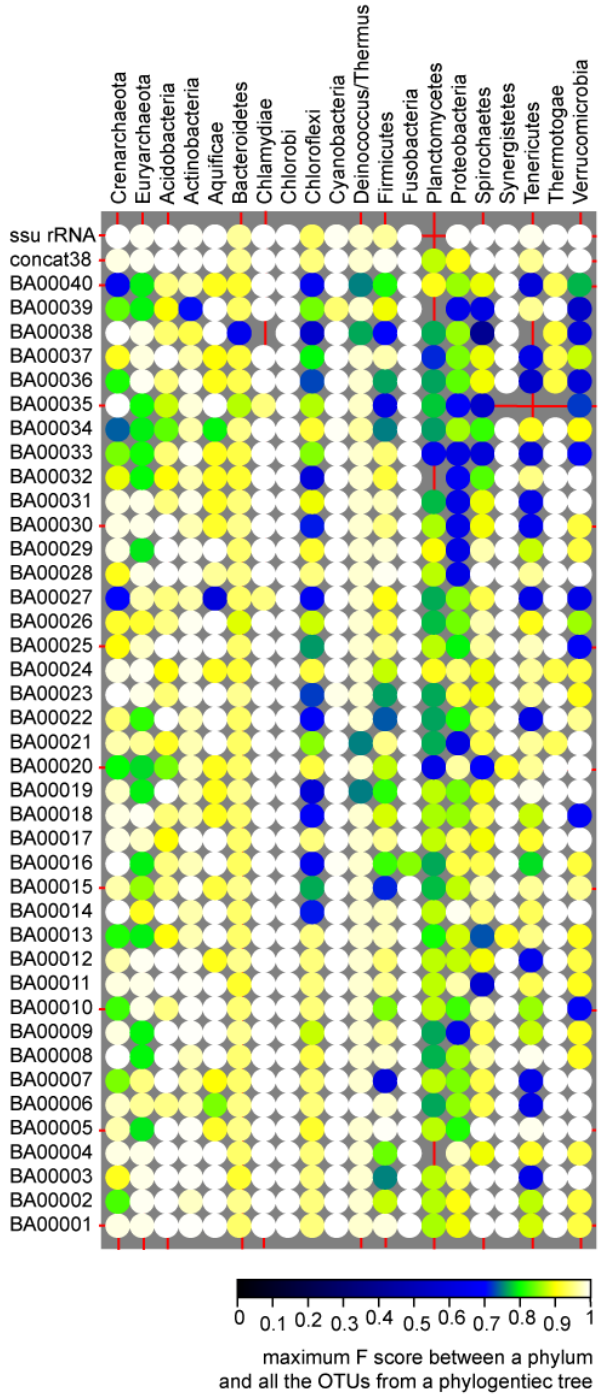

**Figure 6.** The adjusted maximum F1 scores of the phyla in the IMG database comparing with OTUs built by TreeOTU from the phylogenetic trees of ssu-rRNA, 40 PhyEco markers and the concatenated alignment 38 PhyEco markers. Only phyla with more than 10 taxa presented in a given tree are listed.

**Table:**

**Table S1.** Genome numbers and function descriptions of the single-copied PhyEco marker families in the study.

| Marker ID | Genome number | Description |
|---|---|---|
| BA00001 | 4584 | ribosomal protein S2 |
| BA00002 | 4552 | ribosomal protein S10 |
| BA00003 | 4564 | ribosomal protein L1 |
| BA00004 | 4568 | translation elongation factor EF-2 |
| BA00005 | 4575 | translation initiation factor IF-2 |
| BA00006 | 4584 | metalloendopeptidase |
| BA00007 | 4552 | ribosomal protein L22 |
| BA00008 | 4559 | ffh signal recognition particle protein |
| BA00009 | 4542 | ribosomal protein L4/L1e |
| BA00010 | 4559 | ribosomal protein L2 |
| BA00011 | 4608 | ribosomal protein S9 |
| BA00012 | 4568 | ribosomal protein L3 |
| BA00013 | 4501 | phenylalanyl-tRNA synthetase beta subunit |
| BA00014 | 4564 | ribosomal protein L14b/L23e |
| BA00015 | 4588 | ribosomal protein S5 |
| BA00016 | 4528 | ribosomal protein S19 |
| BA00017 | 4554 | ribosomal protein S7 |
| BA00018 | 4570 | ribosomal protein L16/L10E |
| BA00019 | 4500 | ribosomal protein S13 |
| BA00020 | 4552 | phenylalanyl-tRNA synthetase alpha subunit |
| BA00021 | 4605 | ribosomal protein L15 |
| BA00022 | 4552 | ribosomal protein L25/L23 |
| BA00023 | 4598 | ribosomal protein L6 |
| BA00024 | 4540 | ribosomal protein L11 |
| BA00025 | 4586 | ribosomal protein L5 |
| BA00026 | 4529 | ribosomal protein S12/S23 |
| BA00027 | 4529 | ribosomal protein L29 |
| BA00028 | 4573 | ribosomal protein S3 |
| BA00029 | 4570 | ribosomal protein S11 |
| BA00030 | 4597 | ribosomal protein L10 |
| BA00031 | 4586 | ribosomal protein S8 |
| BA00032 | 4491 | tRNA pseudouridine synthase B |
| BA00033 | 4603 | ribosomal protein L18P/L5E |
| BA00034 | 4551 | ribosomal protein S15P/S13e |
| BA00035 | 3562 | Porphobilinogen deaminase |
| BA00036 | 4542 | ribosomal protein S17 |
| BA00037 | 4600 | ribosomal protein L13 |
| BA00038 | 3947 | phosphoribosylformylglycinamidine cyclo-ligase |

| BA00039 | 4458 | ribonuclease HII |
| BA00040 | 4567 | ribosomal protein L24 |